\pdfoutput=1
\documentclass[journal=ancac3,manuscript=article]{achemso}
\usepackage{makeidx}
\usepackage{graphicx}
\usepackage{epstopdf}
\usepackage{subfigure}
\usepackage{color}
\usepackage{float}
\usepackage{amsmath, amssymb}
\usepackage{comment}

\author{Xu Han}
\affiliation{Department of Physics, Hong Kong University of Science and Technology, Hong Kong, China}
\altaffiliation{These authors contributed equally to this work.}
\author{Jiangxiazi Lin}
\affiliation{Department of Physics, Hong Kong University of Science and Technology, Hong Kong, China}
\altaffiliation{These authors contributed equally to this work.}
\author{Junwei Liu}
\affiliation{Department of Physics, Hong Kong University of Science and Technology, Hong Kong, China}
\author{Ning Wang}
\affiliation{Department of Physics, Hong Kong University of Science and Technology, Hong Kong, China}
\author{Ding Pan}
\affiliation{Department of Physics, Hong Kong University of Science and Technology, Hong Kong, China}
\alsoaffiliation{Department of Chemistry, Hong Kong University of Science and Technology, Hong Kong, China}
\email{dingpan@ust.hk}

\title{Effects of Hexagonal Boron Nitride Encapsulation on the Electronic Structure of Few-layer MoS$_2$}

\begin{document}

\begin{abstract}
The hexagonal boron nitride (hBN) encapsulation has been widely used in the electronics applications of 2D materials to improve device performance by protecting 2D materials against contamination and degradation. It is often assumed that hBN layers as a dielectric would not affect the electronic structure of encapsulated 2D materials. Here we studied few-layer MoS$_2$ encapsulated in hBN flakes by using a combination of theoretical and experimental Raman spectroscopy. We found that after the encapsulation the out-of-plane A$_{1g}$ mode is upshifted, while the in-plane E$_{2g}^1$ mode is downshifted. 
The measured downshift of the E$_{2g}^1$ mode does not decrease with increasing the thickness of MoS$_2$,
which can be attributed to tensile strains in bilayer and trilayer MoS$_2$ caused by the typical experimental process of the hBN encapsulation. We estimated the strain magnitude and found that the induced strain may cause the K-Q crossover in the conduction band of few-layer MoS$_2$, so greatly modifies its electronic properties as an n-type semiconductor. 
Our study suggests that the hBN encapsulation should be used with caution, as it may affect the electronic properties of encapsulated few-layer 2D materials. 
\end{abstract}

\section{Introduction}

Few-layer MoS$_2$, a widely studied 2D material, has shown great potential for next-generation electronic devices \cite{splendiani2010emerging,radisavljevic2011single,ganatra2014few}. It can be made into an  n-type \cite{radisavljevic2011single,ganatra2014few} or p-type \cite{zhan2012large} semiconductor with high carrier mobility, and the possible applications range from transistors \cite{radisavljevic2011single} to water splitting electroncatalysts \cite{wan2018engineering}. Its electronic properties can be effectively tuned by the number of stacking layers 
as well as strain. Uniaxial, biaxial, and local strains have been applied, and many interesting phenomena were found \cite{liang2017monitoring,PhysRevB.85.033305}. For example, $\sim$2\% uniaxial strain leads to a direct to indirect gap transition for monolayer MoS$_2$  and $\sim$10\% biaxial strain even converts it to a metal \cite{li2012ideal}.

MoS$_2$ thin films are found to be not stable in air, can be contaminated or oxidized at the surface, so in electronics applications, hexagonal boron nitride (hBN) layers are often used as a corrosion resistant coating to protect MoS$_2$ \cite{cui2015multi,lee2015highly}. The hBN layers can be stable at more than 1000 $^\circ$C in air and oxygen is unable to penetrate through even at high temperatures \cite{liu2013ultrathin,kostoglou2015thermal}.
Moreover, the dangling-bond-free surface of hBN serves as an atomically flat substrate for MoS$_2$, and when the hBN layers are put in between MoS$_2$ and SiO$_2$, they can screen the charge impurities in the SiO$_2$ surface. Thus, the hBN encapsulation greatly improves carrier mobility and channel quality, so that quantum oscillations can be observed \cite{cui2015multi,lin2019determining}.
The hBN encapsulation is now widely used in device applications of many 2D materials, such as transition metal dichalcogenides \cite{ahn2016prevention}, phosphorene \cite{li2017direct}, magic-angle graphene superlattices\cite{cao2018unconventional}.

From monolayer to bulk, the out-of-plane electronic dielectric constant of hBN, $\epsilon_{\infty}^{\bot}$, increases from 2.89 to 3.03, while the in-plane $\epsilon_{\infty}^{\parallel}$ changes little  (4.96$\sim$4.98), according to the first-principles calculation \cite{laturia2018dielectric}.
The band gap of hBN layers is between 5 and 6 eV \cite{zhang2017two}, larger than that of few-layer MoS$_2$. Thus, in 2D electronics applications, hBN layers often work as gate dielectrics for MoS$_2$ \cite{cui2015multi,lee2015highly,zhang2017two}. In the previous studies about interactions between hBN and MoS$_2$, hBN layers were usually treated as substrates, and have been shown to affect the optical properties of MoS$_2$ due to the dielectric screening, such as photoluminescene emission \cite{buscema2014effect}, Raman \cite{buscema2014effect,li2015raman}, and exciton \cite{cadiz2017excitonic} spectra. However, the effects of the hBN encapsulation on the electronic properties of MoS$_2$ have not been well investigated.  
It is often assumed that the band structure and transport properties of MoS$_2$ are not affected by the hBN encapsulation \cite{cui2015multi,lee2015highly}.

Here, by a combination of theoretical and experimental methods, we studied the Raman spectra of few-layer MoS$_2$ encapsulated in hBN flakes. We found that the
typical experimental process of the hBN encapsulation may cause tensile strain in bilayer and trilayer MoS$_2$, while the induced strain in monolayer MoS$_2$ is negligible. The strain due to the hBN encapsulation may change the position of the conduction band minimum of few-layer MoS$_2$, so greatly affects the transport properties of few-layer MoS$_2$ as an n-type semiconductor.

\section{Methods}
\subsection{Experimental methods }
We used the well-developed dry transfer technique to stack the hBN/MoS$_2$/hBN heterostructure \cite{xu2016universal}. First, atomically-thin MoS$_2$ flakes were exfoliated from bulk material onto a silicon wafer with a 280~nm SiO$_2$ layer on top. A monolayer flake was identified using optical contrast. Raman spectroscopy was performed subsequently to this flake to confirm the number of MoS$_2$ layers. Then we prepared two thin hBN flakes, one on another SiO$_2$/Si wafer and the other one on a PMMA film. Using an optical microscope, the hBN layers on PMMA was aligned with the MoS$_2$ flake by the natural cleavage edges of flakes and was used to separate MoS$_2$ from the silicon wafer, then the obtained hBN/MoS$_2$ stack was transferred onto the hBN on the SiO$_2$/Si wafer. Finally, the PMMA film was removed by acetone. The encapsulated hBN/MoS$_2$/hBN heterostructure was annealed at 300 $^\circ$C in an argon protected environment for 6 hours to reduce organic residues and impurities. Raman spectroscopy was performed again to the same MoS$_2$ flake now encapsulated by hBN. Experiments for bilayer and trilayer MoS$_2$ followed the same procedures.

All Raman experiments were performed at ambient conditions using the InVia (Renishaw) micro Raman system with a 514.5~nm laser. The laser power was controlled at $\sim$0.15~mW and $\sim$2.5~mW for the exposed and encapsulated MoS$_2$, respectively, to prevent damage to the sample.

\subsection{DFT Calculations}
Electronic structure calculations were performed using the plane-wave pseudopotential method implemented in the Quantum ESPRESSO package (version 6.1)\cite{giannozzi2017advanced}. The SG15 Optimized Norm-Conserving Vanderbilt (ONCV) pseudopotentials (version 1.1) were used\cite{hamann2013optimized,schlipf2015optimization}.
In the spin-orbital coupling (SOC) calculations, full relativistic pseudopotentials were from Ref \cite{doi:10.1021/acs.jctc.6b00114}. The kinetic energy cutoff for plane waves was 60 Ry. 
The convergence thresholds for energy, force, and stress were 10$^{-5}$ Ry, 10$^{-4}$ Ry/Bohr, and 50 MPa, respectively.
We chose the local density approximation (LDA)\cite{perdew1981self} as the exchange-correlation (xc) functional to calculate Raman frequencies (see below). For the band structure calculations, we used the PBE xc functional \cite{perdew1996generalized} with SOC.
For multiple layers structures, the interlayer distances were obtained by the van der Waals functional optB88-vdW \cite{0953-8984-22-2-022201}. A Monkhorst-Pack k-point mesh of
10$\times$10$\times$1 was used with the primitive cells of few-layer MoS$_2$ and 2$\times$2$\times$1 with the encapsulated MoS$_2$. With periodic boundary conditions, the vacuum between two neighboring images is at least 12 \AA.

The hBN/MoS$_2$/hBN heterostructure was made by 5$\times$5 hBN and 4$\times$4 MoS$_2$. We used three layers hBN to encapsulate MoS$_2$ (see Fig. \ref{1}(b)). The in-plane lattice constant of the supercell was kept the same as that of free-standing MoS$_2$, and we increased the in-plane lattice constant of hBN by 0.8\% to fit the supercell. 

Raman frequencies were calculated by using the frozen phonon method, where the atomic displacement was 0.05 bohr. 
Our phonon frequencies are in very good agreement with the results obtained by density functional perturbation theory \cite{baroni2001phonons}. In the heterostructure calculations, we diagonalized the dynamic matrix $\mathbf{D}(\Vec{q})$ at $\Vec{q}=0$ only:
\begin{equation}
\det|\mathbf{D}(\Vec{q})-\omega^2(\Vec{q})\mathbf{1}|=0,
\end{equation}
where $\Vec{q}$ is the phonon wave vector of MoS$_2$, $w$ is the vibration frequency, and $\mathbf{1}$ is the identity matrix.

\section{Results and Discussion}
\begin{figure}[H]
\centering
\vspace{5mm}
\includegraphics[width=0.6 \textwidth]{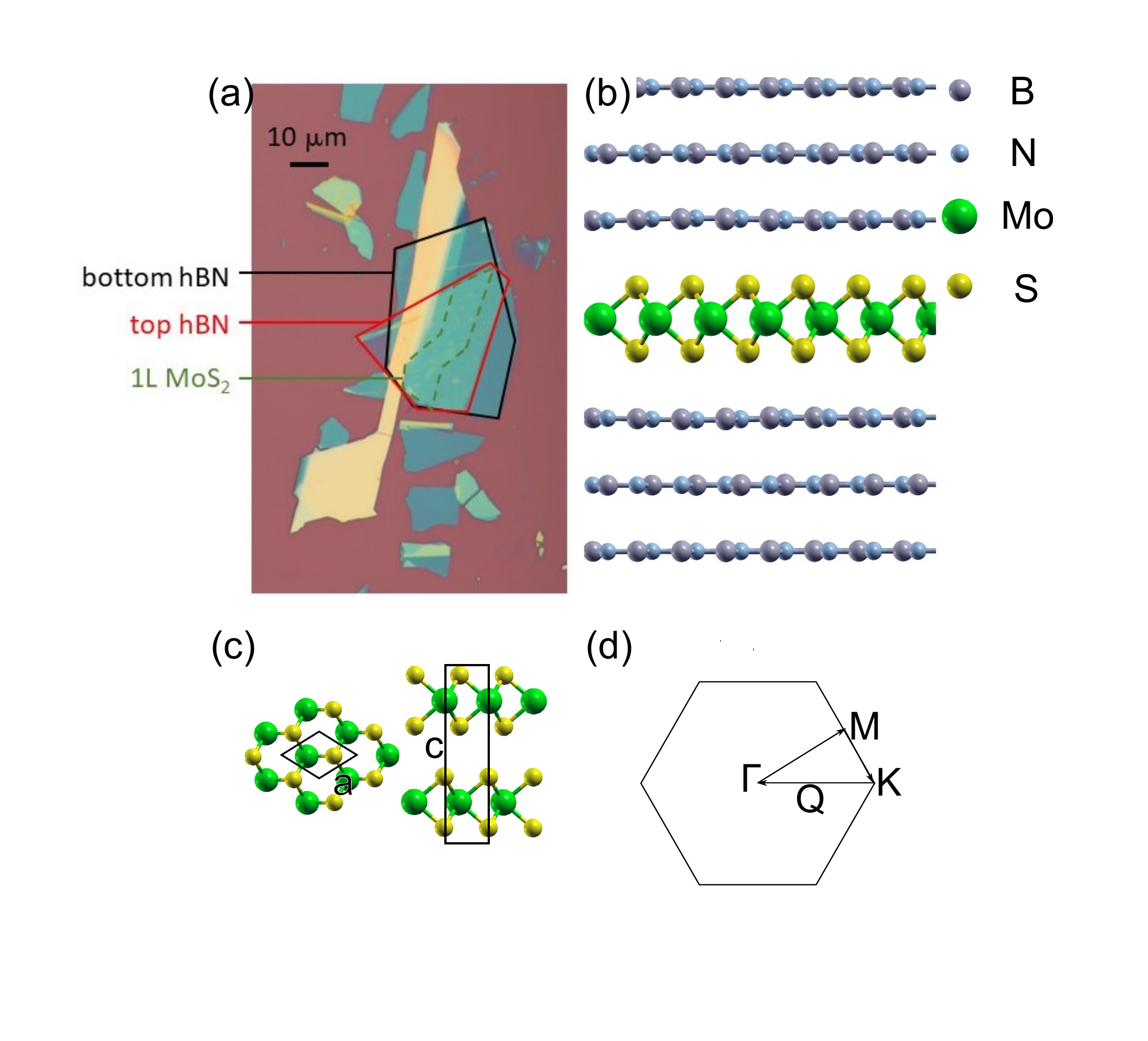}
\caption{Monolayer MoS$_2$ encapsulated in hBN flakes. (a) Optical microscope image of the hBN/MoS$_2$/hBN heterostructure on the SiO$_2$/Si substrate. The bottom hBN, monolayer MoS$_2$, and top hBN flakes are marked by black, green, and red contours, respectively. (b)  Side view of the hBN/MoS$_2$/hBN heterostructure. 
(c) Top and side views of the unit cell of bulk MoS$_2$. (d) First Brillouin zone of MoS$_2$.} 
\label{1}
\end{figure}

\begin{figure}[H]
\centering
\vspace{5mm}
\includegraphics[width=0.6 \textwidth]{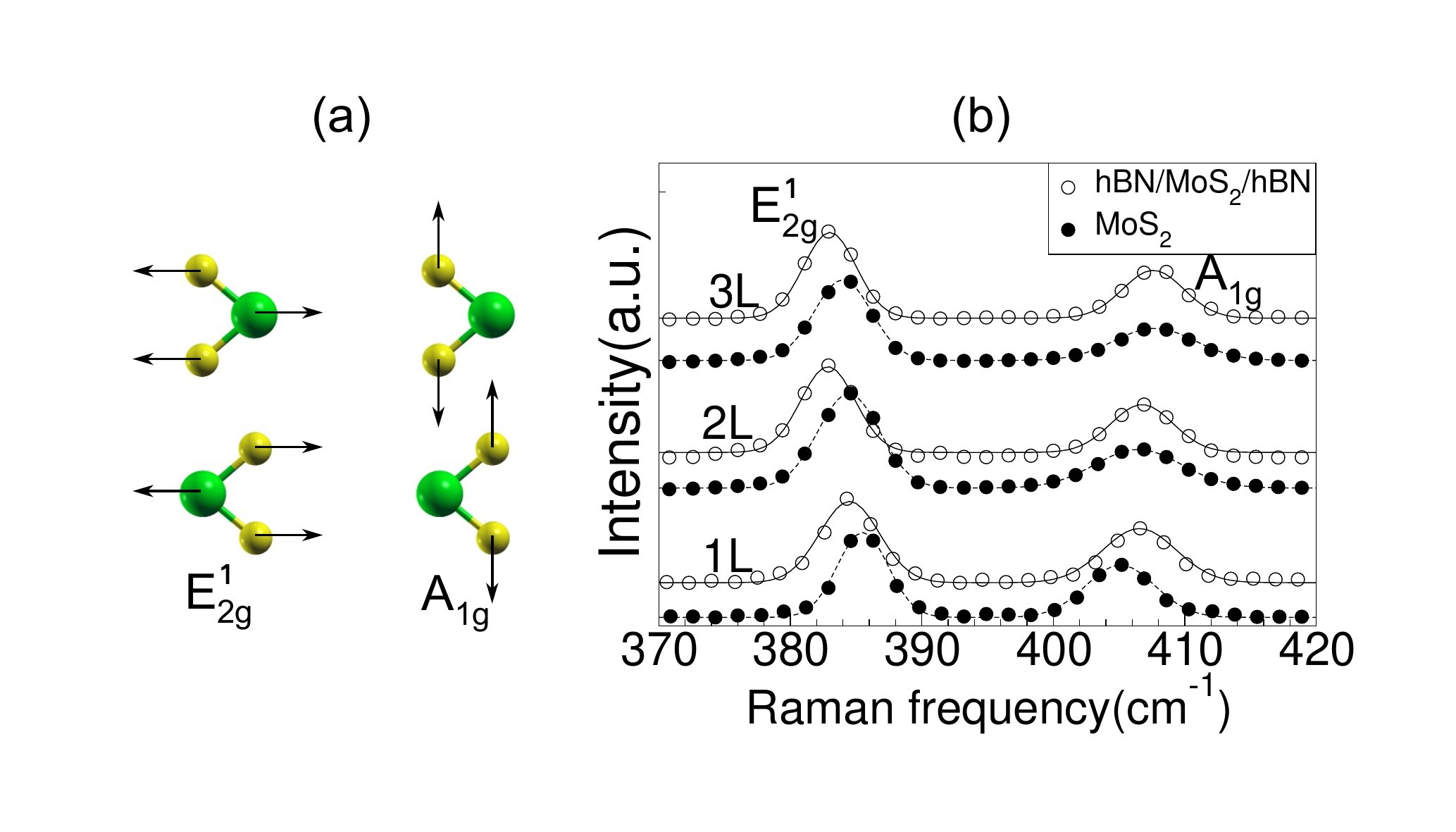}
\caption{Vibrational modes and experimental Raman spectra of MoS$_2$. (a) Atomic displacements of the Raman active modes E$_{2g}^1$ and A$_{1g}$. (b)
Experimental Raman spectra of few-layer MoS$_2$ adsorbed on SiO$_2$/Si (solid dots) and encapsulated in hBN flakes (open circles). Black lines show Gaussian fits. The low-frequency peak corresponds to the E$_{2g}^1$ mode, and the high-frequency peak is for the A$_{1g}$ mode. Monolayer (1L), bilayer (2L), and trilayer (3L) MoS$_2$ are compared.}
\label{Expt_Raman}
\end{figure}

Fig. \ref{1}(a) shows the topview of monolayer MoS$_2$ encapsulated inside two hBN flakes, obtained by optical microscope. In Fig. \ref{Expt_Raman}, we compared the experimental Raman spectra of MoS$_2$ encapsulated in hBN flakes and adsorbed on the SiO$_2$/Si substrate. We increased the thickness of MoS$_2$ from one to three layers to see the change of spectra. Two Raman modes, in-plane (E$_{2g}^1$) and out-of-plane (A$_{1g}$) as shown in Fig. \ref{Expt_Raman}(a), can be seen in the measured spectra in Fig. \ref{Expt_Raman}(b).   
For MoS$_2$ adsorbed on SiO$_2$/Si, with increasing the thickness of MoS$_2$, the frequency of the A$_{1g}$ mode increases, while the E$_{2g}^1$ mode decreases (See Fig. S1(a)), so the frequency difference ($\Delta$) between these two modes becomes larger.
This is why the frequency difference $\Delta$ can be used to count the number of MoS$_2$ layers in experiment \cite{lee2010anomalous}, and our finding is consistent with previous studies \cite{lee2010anomalous, PhysRevB.84.155413}. It has been reported that the stiffening of the out-of-plane mode A$_{1g}$ is attributed to the enhanced interlayer van der Waals (vdW) interactions \cite{lee2010anomalous}, whereas the downshift of the in-plane mode E$_{2g}^1$ is mainly caused by the stronger dielectric screening of the long-range Coulomb interactions \cite{PhysRevB.84.155413}.      

\begin{figure}[H]
\centering
\vspace{5mm}
\includegraphics[width=0.6 \textwidth]{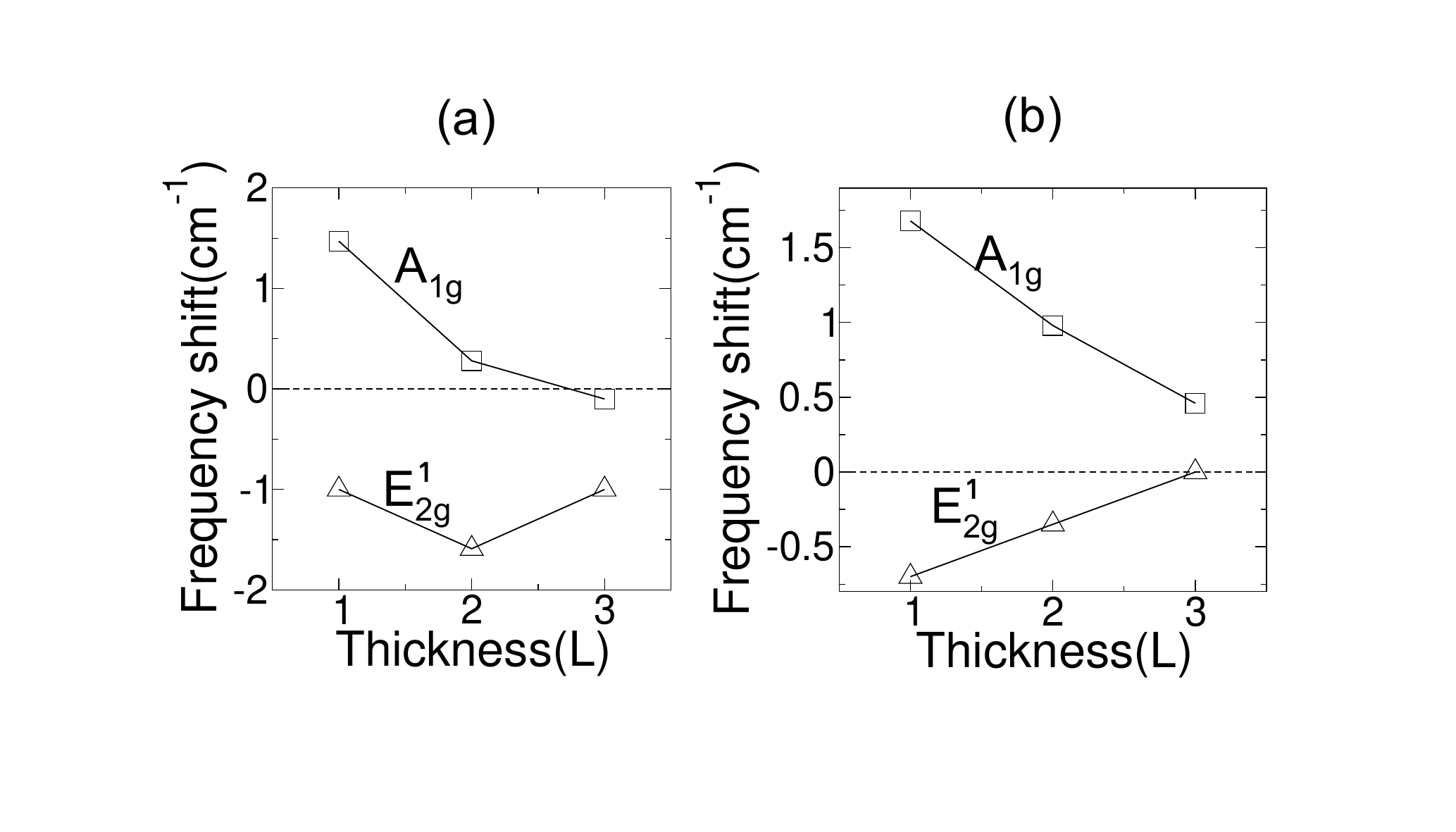}
\caption{Raman frequency shifts as functions of layer thickness. (a) Experimental and (b) calculated Raman frequency shifts ($\delta \omega$) were obtained by comparing the individual Raman modes of the MoS$_2$ layers encapsulated in hBN flakes and adsorbed on the SiO$_2$/Si substrate: 
$\delta \omega= \omega_{hBN} - \omega_{SiO_2/Si}$, where $\omega$ is the Raman frequency of the E$_{2g}^1$ or A$_{1g}$ mode. In the calculations there is zero strain in MoS$_2$.}
\label{Cal_Raman}
\end{figure}

When the MoS$_2$ layers are encapsulated in hBN flakes, the Raman peaks are shifted compared with those obtained from MoS$_2$ on SiO$_2$/Si, as shown in Fig. \ref{Cal_Raman}. The Raman frequency of the out-of-plane mode A$_{1g}$ becomes larger except for trilayer MoS$_2$, while the frequency of the in-plane mode E$_{2g}^1$ decreases, so the frequency difference $\Delta$ becomes larger after the encapsulation. As a result, when we use $\Delta$ to count the number of MoS$_2$ layers inside hBN flakes, caution is needed.
For example, $\Delta$ of monolayer MoS$_2$ in hBN is even larger than that of bilayer MoS$_2$ on SiO$_2$/Si by 0.04 cm$^{-1}$, as shown in Fig. S1 (c). The shift directions of the two modes caused by the hBN encapsulation are the same as the mode shift directions with increasing the thickness of MoS$_2$, so for MoS$_2$ held by hBN, the vdW interactions between MoS$_2$ and hBN layers, and the dielectric screening due to the hBN layers also play important roles in modifying the A$_{1g}$ and E$_{2g}^1$ modes, respectively. 

With increasing the thickness of MoS$_2$ layers, the frequency shifts caused by the hBN encapsulation should become smaller, because the interface effects become less important and the vibration frequencies are getting close to those of bulk MoS$_2$.  In Fig. \ref{Cal_Raman}(a), however, the shift of the E$_{2g}^1$ mode does not decrease with the thickness of MoS$_2$. Instead, bilayer MoS$_2$ has the largest E$_{2g}^1$ mode shift.
Unlike the E$_{2g}^1$ mode, the frequency shift of the A$_{1g}$ mode decreases with thickness, but in trilayer MoS$_2$ it even decreases to a negative value: -0.1 cm$^{-1}$.

To better understand the Raman frequency shifts caused by the hBN encapsulation, we performed density functional theory calculations (see methods). In Table SI, we compared four exchange-correlation(xc) functionals. The semilocal functional PBE \cite{perdew1996generalized} lacks vdW interactions, so it seriously overestimates the interlayer distance in bulk MoS$_2$. When we applied the dispersion correction (PBE-D2)\cite{grimme2006semiempirical} or used the vdW functional (optB88-vdW)\cite{0953-8984-22-2-022201}, the lattice constant $c$ of bulk MoS$_2$ (see Fig. \ref{1}(c)) is improved considerably, but the vibration frequencies are still not as good as the ones obtained using the local density approximation (LDA)\cite{perdew1981self}. Due to the error cancellation, the LDA describes the interlayer interactions remarkably well, so here we used the LDA to compute Raman spectra of few-layer MoS$_2$, as in many previous studies \cite{PhysRevB.84.155413,liang2014first}.

The calculated frequency differences ($\Delta$) between the modes A$_{1g}$ and E$_{2g}^1$ are given in Fig. S1(d). For MoS$_2$ layers adsorbed on SiO$_2$/Si and encapsulated by hBN, the frequency difference $\Delta$ increases with increasing the number of layers, which is consistent with the experimental results in Fig. S1(c). In particular, for one to three MoS$_2$ layers on SiO$_2$/Si, the measured and calculated $\Delta$ values differ within only 0.4 cm$^{-1}$, indicating that our computational settings are very accurate to calculate vibration frequency differences for few-layer MoS$_2$.  
The calculated frequency of the A$_{1g}$ mode of MoS$_2$ in hBN is upshifted compared to that of MoS$_2$ on SiO$_2$/Si, while the E$_{2g}^1$ peak is downshifted, as show in Fig. \ref{Cal_Raman}(b). The shift directions are consistent with those found experimentally; however, the shift magnitudes are different. The experimental downshift of the E$_{2g}^1$ mode is larger than the calculated one, especially for bilayer and trilayer MoS$_2$. In particular, the calculated downshift of the E$_{2g}^1$ mode decreases with increasing the thickness of MoS$_2$ as expected, but the similar trend can not be found in the measured Raman spectra.

The inconsistency between the experimental and calculated Raman data suggests that some other factors may contribute to the measured Raman frequency shifts. Charge transfer and strain are two common reasons. 
Because hBN layers have a very low density of charge impurites, the charge transfer amount is negligible  \cite{buscema2014effect}. Besides, Chakraborty et al. showed that charge transfer affects the A$_{1g}$ mode more than it does E$_{2g}^1$ \cite{PhysRevB.85.161403}, but in our measurements, the shifts of the A$_{1g}$ peak are less obvious than those of E$_{2g}^1$. Thus, we can conclude that charge transfer does not play a major role in our experiments.

\begin{figure}[H]
\centering
\vspace{5mm}
\includegraphics[width=0.6 \textwidth]{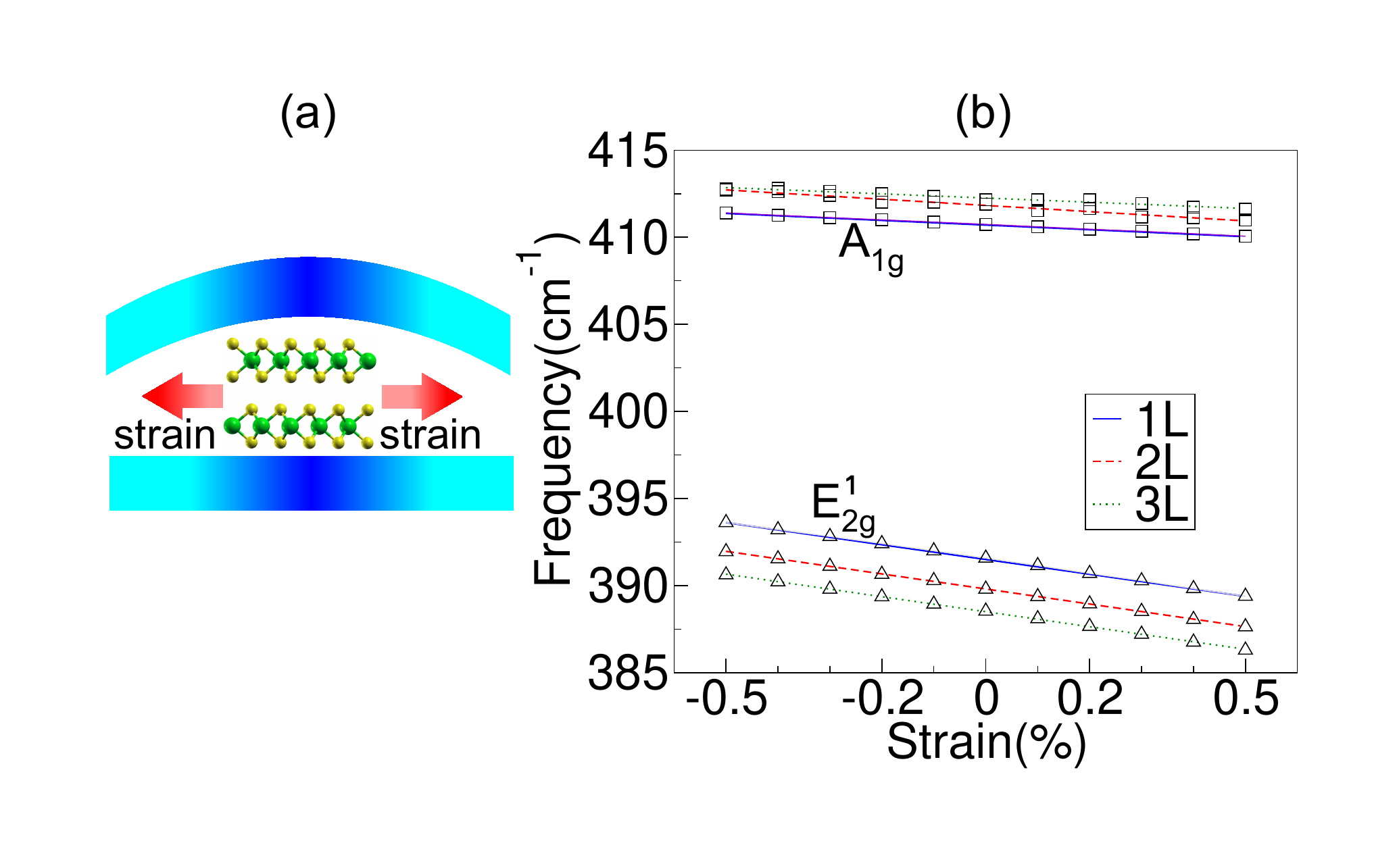}
\caption{Raman frequency vs. biaxial strain. (a) Schematic of the encapsulated bilayer MoS$_2$ with biaxial tensile strain. (b) Calculated Raman frequencies of the E$_{2g}^1$ and A$_{1g}$ modes as functions of biaxial strain. Monolayer (1L), bilayer (2L), and trilayer (3L) MoS$_2$ are compared.}
\label{strain-phonon}
\end{figure}

We consider the strain induced by the hBN encapsulation is biaxial. The biaxial strain is defined as $\epsilon=(a-a_0)/a_0$, where $a$ and $a_0$ are the in-plane lattice constants with and without strain, respectively (see Fig. \ref{1}(c)). Fig. \ref{strain-phonon} shows the vibration frequencies of the A$_{1g}$ and E$_{2g}^1$ modes decrease with increasing the strain of few-layer MoS$_2$. Apparently, the strain affects the E$_{2g}^1$ mode more than A$_{1g}$, so the shift of the E$_{2g}^1$ mode can be used to evaluate the in-plane strain \cite{PhysRevB.87.081307,PhysRevB.93.075401}. 
By polynomial fitting, we found that the frequency of the E$_{2g}^1$ mode has a linear relation with $\epsilon$ in the strain range in Fig. \ref{strain-phonon}. For monolayer,  bilayer, and trilayer MoS$_2$, the $E_{2g}^1$ mode changes by -4.23, -4.23, and -4.31 cm$^{-1}$ per 1$\%$ strain, respectively.
From the difference between the measured and calculated downshifts of the E$_{2g}^1$ mode, we calculated the possible strains, which are 0.06, 0.29, and 0.22\% for monolayer, bilayer, and trilayer MoS$_2$, respectively (see Table I). The tensile strain also causes a tiny downshift of the A$_{1g}$ mode, so the A$_{1g}$ mode frequency of trilayer MoS$_2$ even decreases by -0.1 cm $^{-1}$ after the encapsulation, though the interlayer vdW forces stiffen the A$_{1g}$ mode. 

It is interesting that bilayer and trilayer MoS$_2$ have larger strains than monolayer MoS$_2$.
The lattice mismatch between hBN and MoS$_2$ is as large as 21\%, and the heterostructure layers are stacked together by vdW interactions, so the induced strain is not caused by the lattice mismatch. 
A possible reason is that a thicker MoS$_2$ film might cause a larger deformation of hBN when we heated the heterostructure and pressed the top and bottom hBN layers very hard to squeeze out the air; the MoS$_2$ layers were stretched and could not relax fully when held by the deformed hBN.
We also measured the Raman spectra of the encapsulated bilayer MoS$_2$ before annealing, and found that the biaxial strain is about 0.067\%, much smaller than the strain after annealing, indicating that the experimental annealing process may induce a tensile strain.

Let us see how the electronic structure of MoS$_2$ changes after being encapsulated by hBN.  The band gap of hBN layers is between 5 and 6 eV, which is much larger than those of few-layer or bulk MoS$_2$ indicating that the hBN layers are transparent for MoS$_2$.
In Fig. S2, we unfolded the band structure of the heterostructure supercell using the Brillouin zone of MoS$_2$\cite{PhysRevB.91.041116}
and found that both the valance band top and the conduction band bottom  of the heterostructure come from MoS$_2$, so the semiconductor devices made by hBN/MoS$_2$/hBN heterostructures only show the electronic properties of MoS$_2$.

\begin{figure}[H]
\centering
\vspace{5mm}
\includegraphics[width=0.6 \textwidth]{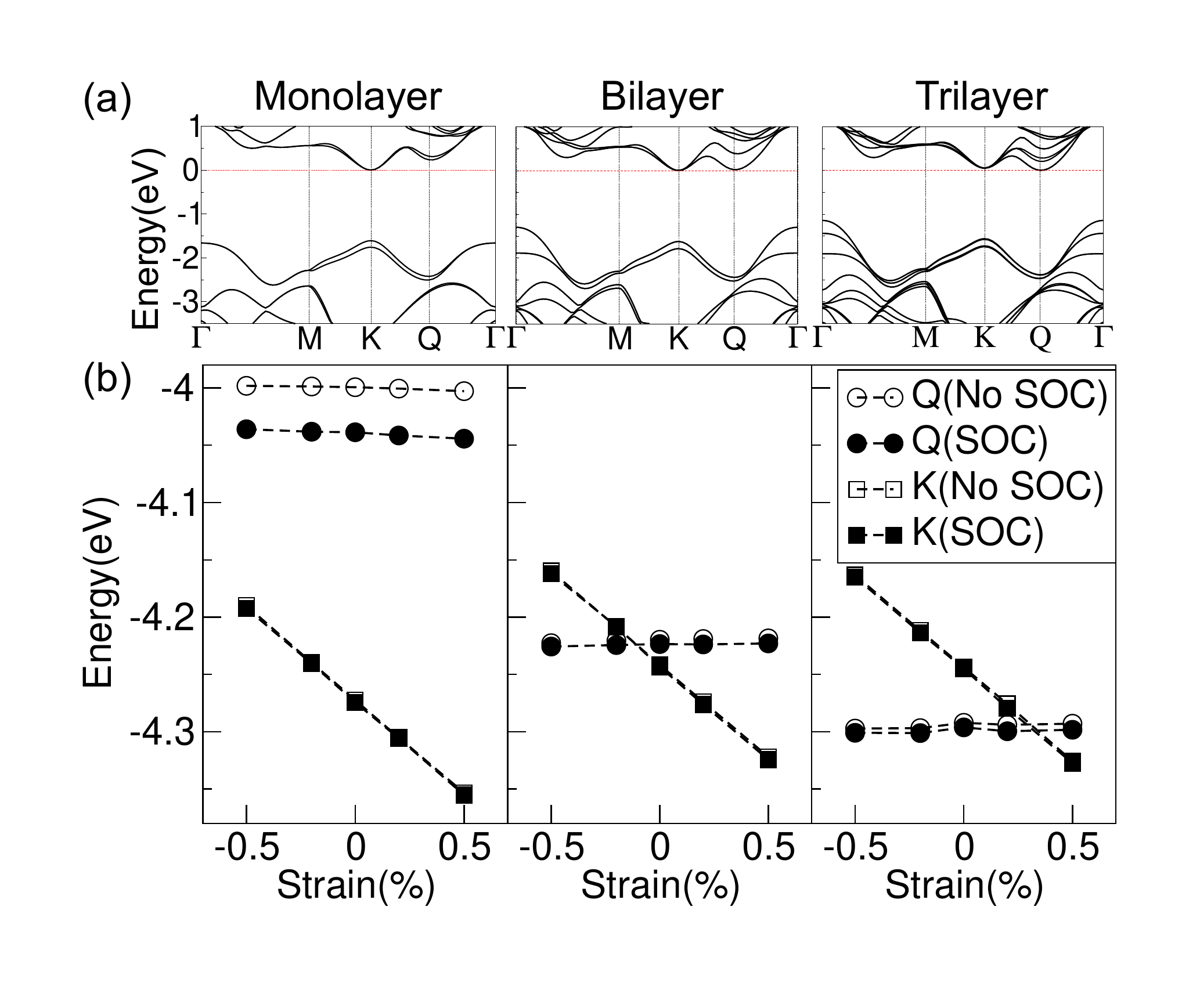}
\caption{Strain effects on the band structure of few-layer MoS$_2$. (a) Band structure of monolayer, bilayer, and trilayer MoS$_2$ at zero strain.  (b)Q (circles) and K (squares) positions in the conduction band with respect to the vacuum level as functions of strain. Calculations with (solid symbols) and without (open symbols) spin-orbital coupling are compared. From left to right: monolayer, bilayer, and trilayer MoS$_2$.}
\label{K-Q}
\end{figure}

The strain induced by the hBN encapsulation affects the electronic properties of MoS$_2$.
Fig. \ref{K-Q} shows the band structures of few-layer MoS$_2$ under strain. 
For bilayer and trilayer MoS$_2$ in the biaxial strain between -0.5\% and +0.5\%, the valance band maximum (VBM) is always at the $\Gamma$ point, whose position moves up with respect to the vacuum level when increasing the biaxial strain. For conduction bands, with increasing the biaxial strain, the K valley position moves down with respect to the vacuum level, whereas the Q point does not change much (see Fig. \ref{K-Q}). Particularly, for the trilayer MoS$_2$, the conduction band minimum (CBM) changes from Q to K when the strain is 0.26\%, which is comparable to the estimated strain caused by the hBN encapsulation. Thus, it is possible that when increasing the thickness of few-layer MoS$_2$, the CBM should move from K to Q, but the tensile strain due to the hBN encapsulation changes it back to the K point.

Since the K and Q valley electrons are very different, the K-Q crossover changes conduction band properties significantly. In the first Brillouin zone of few-layer MoS$_2$,
the valley degeneracy of the K point is twofold, while that of the Q point is sixfold, so the densities of states at these two valley point are different, leading to different Landau level filling factors in quantum oscillation measurements \cite{cui2015multi,lin2019determining}. 
Besides, the effective mass of the Q valley electrons is larger than that of the K valley electrons, as shown in Fig. S4. With increasing the biaxial strain, the effective mass at the Q point increases, while it decreases at the K point, so the hBN encapsulation may amplify the effective mass difference.  
  
\section{Conclusion}
To summarize, we studied few-layer MoS$_2$ encapsulated in hBN flakes by using a combination of density functional theory and experimental Raman spectroscopy. We found that after the encapsulation the out-of-plane A$_{1g}$ mode is upshifted due to the interlayer vdW interactions between hBN and MoS$_2$, while the in-plane E$_{2g}^1$ mode is downshifted, which can be attributed to the dielectric screening of hBN. The measured downshift of the E$_{2g}^1$ mode  does not decrease with increasing the thickness of MoS$_2$, indicating that the typical experimental process of the hBN encapsulation may induce a tensile strain in bilayer and trilayer MoS$_2$.

The strain due to the experimental process of the hBN encapsulation may cause the Q-K crossover in the conduction band of few-layer MoS$_2$. The Q and K valley electrons have different degeneracy and effective masses, so the hBN encapsulation does not only provide a dielectric surrounding for MoS$_2$, but may also substantially affect the transport properties of few-layer MoS$_2$ as an n-type semiconductor. 

The hBN encapsulation has been widely used in many 2D materials applications to improve the device performance and stability. The encapsulation process may also induce the similar tensile strain in those few-layer materials and affect their electronic properties. The combined theoretical and experimental approach introduced here can be used to estimate the magnitude of the strain and to check whether the hBN encapsulation would affect the desired properties of few-layer 2D materials.
The induced strain may be also used to further tune the electronic properties of 
vdW heterostructure devices.

\section{Acknowledgement}
N. W. acknowledges support from Hong Kong Research Grants Council (Project No. GRF-16300717). J. L. acknowledges support from the Hong Kong Research Grants Council (Project No. ECS-26302118). D. P. acknowledges support from Hong Kong University of Science and Technology by the start-up grant and from the Croucher Foundation through the Croucher Innovation Grant.

\begin{suppinfo}
\begin{itemize}
  \item Table SI: Calculated lattice constants of bulk MoS$_2$ and Raman frequencies of the E$_{2g}^1$ and A$_{1g}$ modes in monolayer MoS$_2$.
  \item Figure S1: Calculated Raman spectra of few-layer MoS$_2$ adsorbed on SiO$_2$/Si and encapsulated in hBN flakes. 
  \item Figure S2: Calculated band structure of monolayer MoS$_2$ with and without the hBN encapsulation.
  \item Figure S3: Calculated band structure of few-layer MoS$_2$ under strain.
  \item Figure S4: Calculated effective masses of the Q and K valley electrons in few-layer MoS$_2$.
\end{itemize}
This information is available free of charge via the Internet at http://pubs.acs.org
\end{suppinfo}

\bibliography{ref}

\newpage

\begin{table}
\centering
\caption{Calculated biaxial strain induced by the hBN encapsulation in monolayer (1L), bilayer (2L), and trilayer (3L) MoS$_2$. The uncertainties are obtained from linear regression errors.}
\label{table 1}
\begin{tabular}{c|ccc}
\hline
\hline
&1L&2L&3L\\
\hline
strain(\%)&0.0616$\pm$0.0005&0.2882$\pm$0.0015&0.2231$\pm$0.0015\\
\hline
\end{tabular}
\end{table}

\end{document}